\documentclass[12pt,a4paper]{article}

\usepackage{subfigure,graphicx}
\usepackage{amsmath,amsfonts,latexsym,amssymb,euscript,xr}

\usepackage{authblk}
\usepackage{graphicx}
\usepackage{url}

\begin{document}

\title{\large $\ $\\ \bf \texttt{cyTRON} and \texttt{cyTRON/JS}: two Cytoscape-based applications for the inference of cancer evolution models}

\author[1]{Lucrezia Patruno\footnote{Equal contributors.}}
\author[2,3]{Edoardo Galimberti$^{\ast}$}
\author[4]{Daniele Ramazzotti}
\author[5]{Giulio Caravagna}
\author[1]{Luca De Sano}
\author[1,6]{Marco Antoniotti}
\author[1,7]{Alex Graudenzi}
\affil[1]{Department of Informatics, Systems and Communication, University of Milano-Bicocca, Milan, Italy}
\affil[2]{Department of Computer Science, University of Turin, Turin, Italy}
\affil[3]{ISI Foundation, Turin, Italy}
\affil[4]{Department of Pathology, Stanford University, Stanford, CA, United States}
\affil[5]{The Institute of Cancer Research - ICR, London, UK}
\affil[6]{Milan Center for Neuroscience, Milan, Italy}
\affil[7]{Institute of Molecular Bioimaging and Physiology - IBFM of the National Research Council - CNR, Segrate, Milan, Italy}

\date{}

\maketitle

\begin{abstract}
The increasing availability of sequencing data of cancer samples is fueling the development of algorithmic strategies to investigate tumor heterogeneity and infer reliable models of cancer evolution. We here build up on previous works on cancer progression inference from genomic alteration data, to deliver two distinct \texttt{Cytoscape}-based  applications, which allow to produce, visualize and manipulate cancer evolution models, also by interacting with public genomic and proteomics databases. In particular, we here introduce \texttt{cyTRON}, a stand-alone \texttt{Cytoscape} app, and \texttt{cyTRON/JS}, a web application which employs the functionalities of \texttt{Cytoscape/JS}. 

cyTRON was developed in Java; the code is available at \\ https://github.com/BIMIB-DISCo/cyTRON and on the Cytoscape App Store http://apps.cytoscape.org/apps/cytron. 
cyTRON/JS was developed in JavaScript and R; the source code of the tool is available at https://github.com/BIMIB-DISCo/cyTRON-js and the tool is accessible from https://bimib.disco.unimib.it/cytronjs/welcome.
\end{abstract}

\section{\bf Scientific Background}

Cancer is a complex disease, whose development is caused by the accumulation of alterations in the genome. Some alterations may confer a selective advantage to cancer cells, and this may result in the expansion of cancer clones. In order to understand how cancer evolves, it is of great importance to understand how such \emph{driver} alterations accumulate over time \cite{nowell1976clonal,burrell2013causes}. This goal can be pursued by reconstructing cancer evolution models, which are graphs that encode the evolutionary history of drivers and their temporal relationships. The reconstruction of such models is a complex task mainly because of two reasons: first, much of the data publicly available from initiatives such as \texttt{TCGA} [\url{https://portal.gdc.cancer.gov/}] comes from cross-sectional samples, and hence they lack of temporal information. The second main reason can be found in the heterogeneity of tumors \cite{PicNic,ramazzotti2018modeling}. 

\section{\bf Materials and Methods}

In order to learn meaningful evolution models, we developed a pipeline, \texttt{PICNIC} \cite{PicNic}, which includes the following steps: $i)$ identification of homogeneous sample subgroups (e.g., tumor subtypes), $ii)$ identification of drivers (i.e., the nodes of the output model), $iii)$ identification of mutually exclusive patterns, $iv$) inference of cancer evolution models via distinct algorithms (e.g., \cite{loohuis2014inferring,Capri,Trait}). This pipeline was implemented within the widely used \texttt{TRONCO} \texttt{R} suite for TRanslational ONCOlogy \cite{tronco}, \cite{antoniotti2015design}, which was recently employed, for instance, to analyze the largest kidney cancer cohort currently available \cite{turajlic2018deterministic}.

\begin{figure}[t]
    \centering
	\includegraphics[width=1.0\textwidth]{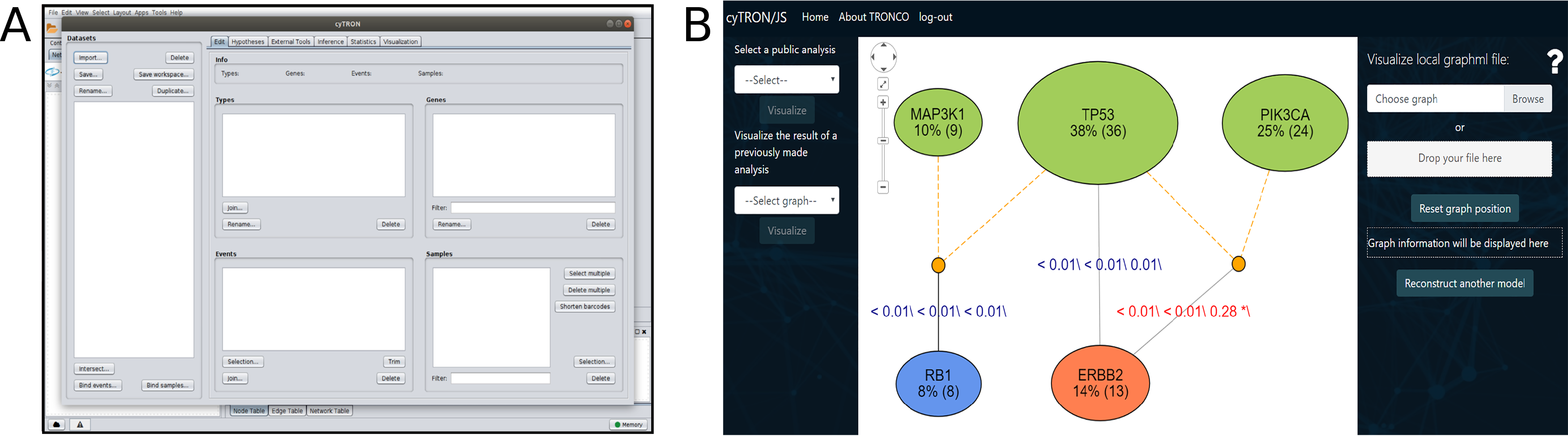}
	\caption{We show in the figure a view of \texttt{cyTRON} workspace (A) and an example of output model by \texttt{cyTRON/JS} (B).}
	\label{fig:output}
\end{figure}

However, \texttt{TRONCO} presents two practical limitations: first, it requires at least some basic programming skills due to its underlying \texttt{R} infrastructure; second, \texttt{TRONCO} is not integrated with publicly available genomic databases, hence providing a non-interactive visualization of the output graphs. 

Therefore, to improve the practicality, effectiveness, interactivity and diffusion of our framework, we integrated it within \texttt{Cytoscape}, an user-friendly open-source platform for the visualization and manipulation of complex networks \cite{shannon2003cytoscape}. We here present \texttt{cyTRON}, a stand-alone \texttt{Cytoscape} app, and \texttt{cyTRON/JS} a web application which employs the functionalities of \texttt{Cytoscape/JS}, both of which allow to produce, visualize and manipulate cancer evolution models, also by interacting with public genomic databases. Figure \ref{fig:output}B shows an example of the output in \texttt{cyTRON/JS}, which exploits Cytoscape/JS to provide an interactive visualization of the evolution model. 

\texttt{cyTRON} and \texttt{cyTRON/JS} were designed for two main purposes: 
\begin{itemize}
	\item Providing an \emph{interactive} and \emph{user-friendly} visualization of TRONCO models: while \texttt{TRONCO} \texttt{R}-based graph display is static, \texttt{cyTRON} and \texttt{cyTRON/JS} provide interactive views, which allow to directly retrieve information about genes involved in the study, by accessing widely-used public genome databases. 
	\item Making TRONCO  \emph{accessible} to users unfamiliar with R programming: \texttt{cyTRON} and \texttt{cyTRON/JS} provide interfaces that enable the usage of \texttt{TRONCO} respectively from \texttt{Cytoscape} and a Web browser, thus removing the need for users to execute any code in order to complete a whole analysis. 
\end{itemize}

The architecture of both tools can be conceptually defined as follows: 
\begin{itemize}
	\item The \emph{front-end} side is composed of an interface that can be used to:
	\begin{enumerate}
		\item Select input data for the \texttt{TRONCO} analysis \cite{tronco}: the input files should be either \texttt{MAF}, \texttt{GISTIC} or user-defined Boolean matrices that contain information about the mutations observed in each sample. 
		\item Set the parameters for the inference in order to access most \texttt{TRONCO} capabilities. Users can specify which driver mutations to include in the analysis, which algorithm among those implemented in \texttt{TRONCO} to use for the reconstruction and the algorithm's corresponding parameters. 
		\item Visualize cancer evolution models and dynamically interact with the result: for instance, by clicking on the genes of the output graph, it is possible to retrieve the information available on public genomic databases. \texttt{cyTRON} gives access to gene information through databases such as \texttt{Ensembl}\footnote{\url{http://www.ensembl.org/index.html}} and \texttt{Entrez}\footnote{\url{https://www.ncbi.nlm.nih.gov/search/}}, that are accessible through the Cytoscape interface. 
		\newline In \texttt{cyTRON/JS} the data displayed for each node are retrieved from the Entrez database \texttt{Gene} \footnote{\url{https://www.ncbi.nlm.nih.gov/gene/}} using \texttt{E-Utils}, an API provided by the National Center for Biotechnology Information. 
	\end{enumerate} 
	\item The \emph{back-end} side includes the communication channel with R. For \texttt{cyTRON}, a Java bridge with R is built by means of \texttt{rJava}. Instead, for \texttt{cyTRON/JS} it is based on \texttt{js-call-r}, a \texttt{Node.js} package which collects the data and parameters set by the user, encodes them in \texttt{JSON} and sends them to R. Then, R commands are transparently executed in order to perform any specific step of the analysis by TRONCO. 
\end{itemize}

Figure \ref{fig:output} shows a view of \texttt{cyTRON} workspace (left) and an example of output model by \texttt{cyTRON/JS} (right). In order to choose between the two tools, users should take into consideration the data and the type of analysis they need to carry out. In particular, since \texttt{cyTRON/JS} is a web application, it is readily accessible from any device and computations are carried out on the back-end side. This feature is useful in case a user needs to carry out a computational-expensive analysis. However, \texttt{cyTRON} is more complete with respect to all the functionalities implemented in TRONCO: for example, it implements also the option of testing hypothesis on mutations through the algorithm Capri \cite{Capri}. 

\section{\bf Conclusion and future work}

\texttt{TRONCO} is an \texttt{R} package that implements state-of-the-art algorithms for the inference of cancer evolution models with the ultimate goal of understanding the evolutionary trajectories driving tumor evolution. In such a multidisciplinary domain, where computer scientists actively cooperate with biologists, being capable of visually understanding the data is crucial to both parties. In order to effectively allow the use of \texttt{TRONCO}, we here presented \texttt{cyTRON} and \texttt{cyTRON/JS}, two \texttt{Cytoscape}-based applications which translate many of \texttt{TRONCO} functionalities into \texttt{Cytoscape}. Our effort aims at designing user-friendly and accessible tools to support the user in the task of exploring cancer genomic data. 

As the TRONCO functionalities are constantly updated and improved, these two new tools need to be kept up to date. Thus, future work will be focused on integrating TRONCO's new algorithms for analyzing single cell datasets \cite{Trait}. In addition to this, \texttt{cyTRON/JS} needs to be extended with the hypothesis testing functionality, in order to enable users to carry out more complex analysis. 

\section*{\bf Acknowledgments}

\texttt{cyTRON} and \texttt{cyTRON/JS} were developed within the Google summer of Code program, in collaboration with the NRNB organization. 
The authors declare no conflicts of interest. 

\vspace{0.5cm}
\bibliographystyle{apalike}
\bibliography{bibliography}

\end{document}